# Thermodynamics of dilute XX chain in a field

P.N. Timonin

*Physics Research Institute at Southern Federal University,
344090, Stachki 194, Rostov-on-Don, Russia, pntim@live.ru*

The isotropic spin one-half XY chain in transverse field (XX chain) has specific ground state phase with permanent criticality (quasi-long-range ordered, QLRO, phase) which exists in a field less than nearest-neighbor exchange. It is characterized by gapless excitations and power law decay of transverse correlators. The dilution of XX chain has drastic effect on this phase – the infinite series of quantum phase transitions marked by magnetization jumps appears in it. The thermodynamics of dilute XX chain allows exact analytical description revealing the peculiarities of the appearing transitions. We calculate the low-temperature magnetization, entropy, longitudinal magnetic susceptibility and specific heat of dilute model to elucidate the influence of quantum transitions on their field and low-temperature behavior. The changes of pair spin correlators under dilution are also analyzed. We argue that other dilute quantum spin chains and ladders with the gapless (algebraic) spin-liquid states would also exhibit the quantum jumps under variation of couplings and/or field similar to those of XX chain.

## 1. Introduction

The discovery of equivalence of *XY* spin-chains with spin one-half to the free fermions [1] was a breakthrough in the studies of quantum phase transitions. Such transition takes place in *XY* chain at *T* = 0 under variation of transverse field *H* when its module becomes equal to the module of the average nearest-neighbor exchange $J = (J_x + J_y)/2, \ J_x J_y > 0$ [1-5]. Usually it features the ordinary scaling behavior with order parameter being the magnetization's component with largest exchange, i. e. $M_x$ if $|J_x| > |J_y|$. Then $M_x$ vanishes at $|H| > |J|$ and here only $M_z$ along the field exists.

Apparently, this scenario does not hold in a special case of isotropic (XX) chain with $J_x = J_y = J$ where rotational symmetry and low dimension makes $M_x = M_y = 0$ at all *H* [1-5]. Nevertheless XX chain also experiences ground state quantum transition at $|H| = |J|$ from saturated phase with $M_z = 1/2 \, sign(H)$ into so-called quasi-long-range-ordered (QLRO) phase at $|H| < |J|$ characterized by the vanishing a gap between ground and excited states in the energy spectrum of free fermions [5-7] and power law decay of spin correlators [8-10]. Thus XX chain is always in critical state at all $|H| < |J|$ while anisotropic *XY* chain is gapless only at transition point.

This specificity of isotropic chain can be easily understood noting that in the fermionic language QLRO is a "metal" phase while the saturated one at $|H|>|J|$ is an "isolator". Indeed, the one-fermion spectrum of this chain is [1-4]

$$\varepsilon(q) = J(\cos\pi q - h), \quad h = H/J, \qquad (1)$$

where $q$ is a continuous wave-number in the interval $0 \leq q \leq 1$ for infinite chain. Here number of fermions is not conserved so both chemical potential $\mu$ and Fermi energy $\varepsilon_F$ are zero, $\mu = \varepsilon_F = 0$. Thus at $|h|<1$ Fermi energy lies within a "conduction band" and spectrum is gapless. Under $h$ variation the state of this "metal" changes continuously with the changes of the Fermi wave-number

$$q_F(h) = \frac{1}{\pi}\arccos h, \qquad (2)$$

i.e. $q_F(0) = 1/2$ so we have half-filled band while $q_F(1) = 0$ and band is fully filled or empty depending on the $J$ sign.

The further insight in the nature of this permanent criticality in the QLRO phase is given by the studies of ground states of finite XX chains [6, 7]. In finite chains the critical state splits into a series of quantum transitions due to discrete (rational) values of $q$ in the same dispersion law (1). The transitions take place when $q_F(h)$ becomes equal to one of allowed rational $q$. At these points the ground state is changed via adding or deleting of fermion with wave-number nearest to $q_F(h)$. Right at the transition point ground state is doubly degenerate as both states with and without fermion with $\varepsilon_F = 0$ have the same energy. Thus the level crossing in the energy spectrum constitutes the mechanism of quantum phase transitions in XX chain [6, 7] and QLRO phase in the infinite chain limit is a consequence of level crossing at every $q$.

In real crystals with non-interacting XX chains the detection of QLRO phase at low temperature $T$ can be rather complicated as the only pronounced feature of macroscopic chains is power law decay of correlators [8-10]. Yet the manifestations of this phase can be more visual in dilute crystals where non-magnetic ions substitute magnetic ones in the chains. Then discrete level crossing transitions in finite magnetic segments can be seen in usual macroscopic experiments at low $T$ as anomalies in field dependencies of magnetization, specific heat etc. So one should have some quantitative predictions on these anomalies to recognize the presence of XX chains in real crystals. Here we present the study of macroscopic thermodynamics of dilute XX chains showing the spectacular features of QLRO phase at low $T$.

## 2. Thermodynamics of QLRO phase

First we recall the fermionization procedure for finite XX chain with spin one-half [1, 6, 7]. We start with the Hamiltonian for *N* spins with free boundaries

$$\mathcal{H}_N / J = \frac{1}{4}\sum_{n=1}^{N-1}\left(\sigma_n^x \sigma_{n+1}^x + \sigma_n^y \sigma_{n+1}^y\right) - \frac{h}{2}\sum_{n=1}^{N}\sigma_n^z,$$

where $\sigma_n^\alpha$, $\alpha = x, y, z$ are Pauli matrices defining operators of local magnetic moments

$$s_n^\alpha = \sigma_n^\alpha / 2.$$

Introducing the fermionic creation and annihilation operators through Jordan-Wigner transformation

$$a_n^+ = \sigma_n^+ \prod_{i=1}^{n-1}\left(-\sigma_i^z\right), \quad a_n = \sigma_n^- \prod_{i=1}^{n-1}\left(-\sigma_i^z\right), \quad \sigma_n^\pm = \left(\sigma_n^x \pm i\sigma_n^y\right)/2 \quad (3)$$

and using the relation

$$\sigma_n^z = 2\sigma_n^+ \sigma_n^- - 1 = 2a_n^+ a_n^- - 1 \quad (4)$$

we get

$$\mathcal{H}_N / J = \frac{1}{2}\sum_{n=1}^{N-1}\left(a_n^+ a_{n+1} + a_{n+1}^+ a_n\right) - h\sum_{n=1}^{N} a_n^+ a_n + \frac{Nh}{2}$$

This Hamiltonian is diagonalized via transformation to new fermionic operators $b_k$

$$a_n = \sqrt{\frac{2}{N+1}}\sum_{k=1}^{N} b_k \sin\frac{\pi k n}{N+1} \quad (5)$$

In terms of $b_k$ and $b_k^+$ we have the Hamiltonian of free fermions,

$$\mathcal{H}_N / J = \sum_{k=1}^{N}\left(\cos\frac{\pi k}{N+1} - h\right) b_k^+ b_k + \frac{Nh}{2}$$

With this simple expression one can easily obtain thermodynamic potential of dilute XX chain being the sum of independent contributions from finite fragments appearing under dilution. Therefore the average potential per site in case of random dilution is

$$F = -T\sum_{l=1}^{\infty}\frac{N_l}{N}\ln Tr\exp\left(-\frac{\mathcal{H}_l}{T}\right)$$

Here $N_l$ is the average number of magnetic clusters with $l$ sites. If magnetic ions appear independently on all sites with probability $p$ then for large $N$ [11]

$$N_l \approx N(1-p)^2 p^l$$

so in the limit $N \to \infty$

$$F = -T\sum_{l=1}^{\infty}(1-p)^2 p^l \ln Tr\exp(-\beta \mathcal{H}_l) = -T(1-p)^2 \sum_{l=1}^{\infty} p^l \sum_{k=1}^{l} \ln\left(1+e^{2u_{k,l}}\right) + phJ/2 \quad (6)$$

Here

$$u_{k,l} = \frac{J}{2T}\left[h - \cos\left(\frac{\pi k}{l+1}\right)\right] \quad (7)$$

The thermodynamic potential in Eq. (6) allows to find macroscopic observables of dilute XX chain averaged over disorder. Thus average magnetization along the field is

$$M = \langle\langle\sigma^z\rangle_T\rangle_p / 2 = -\frac{1}{J}\frac{\partial F}{\partial h} = (1-p)^2 \sum_{l=1}^{\infty} p^l \sum_{k=1}^{l}\left(1+e^{-2u_{k,l}}\right)^{-1} - p/2$$

$$= \frac{(1-p)^2}{2}\sum_{l=1}^{\infty} p^l \sum_{k=1}^{l} \tanh u_{k,l} \quad (8)$$

Here $\langle\langle...\rangle_T\rangle_p$ is shorthand notation for Gibbs average followed by the average over possible realizations of disorder.

Similarly we get the following expressions for the longitudinal magnetic susceptibility

$$\chi_{zz} \equiv \chi = \frac{\partial M}{J\partial h}, \quad \chi = \frac{(1-p)^2}{4T}\sum_{l=1}^{\infty} p^l \sum_{k=1}^{l}\cosh^{-2}(u_{k,l}) \quad (9)$$

entropy $S$ and specific heat $C$,

$$S = -\frac{\partial F}{\partial T} = (1-p)^2 \sum_{l=1}^{\infty} p^l \sum_{k=1}^{l}\left[\ln(2\cosh u_{k,l}) - u_{k,l}\tanh u_{k,l}\right] \quad (10)$$

$$C = T\frac{\partial S}{\partial T} = \frac{(1-p)^2}{2}\sum_{l=1}^{\infty} p^l \sum_{k=1}^{l} u_{k,l}^2 \cosh^{-2}(u_{k,l}) \quad (11)$$

At $p \to 1$, Eqs. (6, 8-11) become the corresponding expressions for pure infinite chain. To see this we note that in this limit the major contribution to the sum over $k$ transforms into integral as

$$(1-p)^2 \sum_{l=100}^{\infty} p^l \sum_{k=1}^{l} X\left(\pi \frac{k}{l+1}\right) \to (1-p)^2 \sum_{l=100}^{\infty} p^l (l+1) \int_0^\pi \frac{d\theta}{\pi} X(\theta) \to \int_0^\pi \frac{d\theta}{\pi} X(\theta).$$

For example, at $p=1$ we get the known expression for magnetization of infinite XX chain [2-4]

$$M = \int_0^\pi \frac{d\theta}{2\pi} \tanh \frac{J}{2T}(h - \cos\theta). \qquad (12)$$

The above expressions for thermodynamic quantities allow only approximate calculations of them at $T > 0$ as one can calculate only finite sums. As the $l$-th term of the series (8) - (11) is of the order $lp^l$ the finite sums with $l < L$ give approximate values with relative error $\varepsilon$ which we can roughly estimate as

$$\varepsilon \approx \sum_{l=L+1}^{\infty} lp^l \Big/ \sum_{l=1}^{\infty} lp^l = p^L \left[L(1-p)+1\right]. \qquad (13)$$

Thus to achieve, e.g. $\varepsilon$ = 0.05 for $p$ = 0.25, 0.5, 0.75 we should take $L$ = 3, 7, 16 terms in Eqs. (8-11) correspondingly. The results of such calculations with $\varepsilon < 0.02$ ($L$ = 20) are shown in Fig. 1. Here field dependencies of $M$, $\chi$, $S$ and $C$ at $T = 0.1J$ feature somewhat smeared anomalies reminiscent of ground state quantum transitions in finite magnetic clusters.

Actually at $T$ = 0 dilute XX chain has infinite set of these anomalies at all fields for which $q_F(h)$ is rational. Let us consider field dependency of $M$ at $T$ = 0. Here we get from (6)

$$2M = (1-p)^2 \sum_{l=1}^{\infty} p^l \sum_{k=1}^{l} sign\left[k - q_F(h)(l+1)\right], \quad sign(0) = 0$$

Summation over $k$ then gives

$$2M = (1-p)^2 \sum_{l=1}^{\infty} p^l \left\{l - 2m_l^F + \delta_{m_l^F, q_F(h)(l+1)}\right\} \qquad (14)$$

Here $m_l^F \equiv \left[q_F(l+1)\right]$ is an integer part of $q_F(l+1)$ and Kronecker delta is 1 when $q_F(h)(l+1)$ is integer and 0 otherwise.

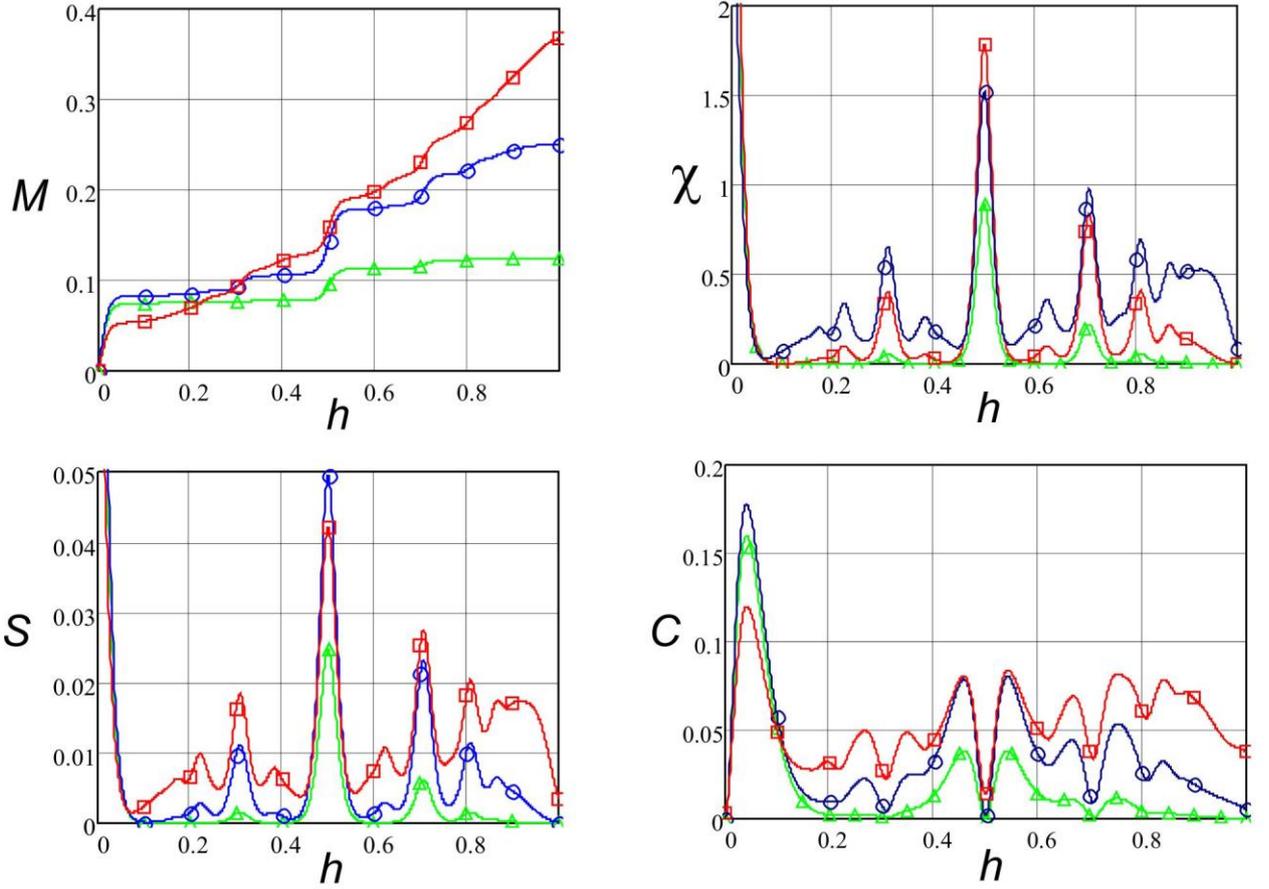

*Fig.1. Field dependencies of magnetization M, susceptibility along the field $\chi$, entropy S and specific heat C of dilute XX chain at T=0.1J for p = 0.25 (△), 0.5(o), 0.75(□). The sums in Eqs. (8-11) are limited by L = 20 terms.*

Apparently M in (14) has jumps at every rational $q_F(h)$. This means that in every field interval there are infinitely many jumps. However, most of them are tiny and unobservable in real experiments. The approximate M with largest jumps at $h < 1$ is shown in Fig. 2 for p = 0.25, 0.5, 0.75 as compared with p = 1. Note that according to (12) at T = 0, p=1, h < 1 we have the known result [2-4]

$$M = \frac{1}{2} - q_F(h) \qquad (15)$$

while the data for p < 1 are obtained from (14) reducing the sum by l < 20. As Fig. 2 shows, the noticeable feature of dilute chain's magnetization is that it can be larger than that of pure chain at the same field. Qualitatively this effect is the consequence of a freedom of finite magnetic fragments to orient their magnetization along the field which can overcome the diminishing of overall M due to dilution.

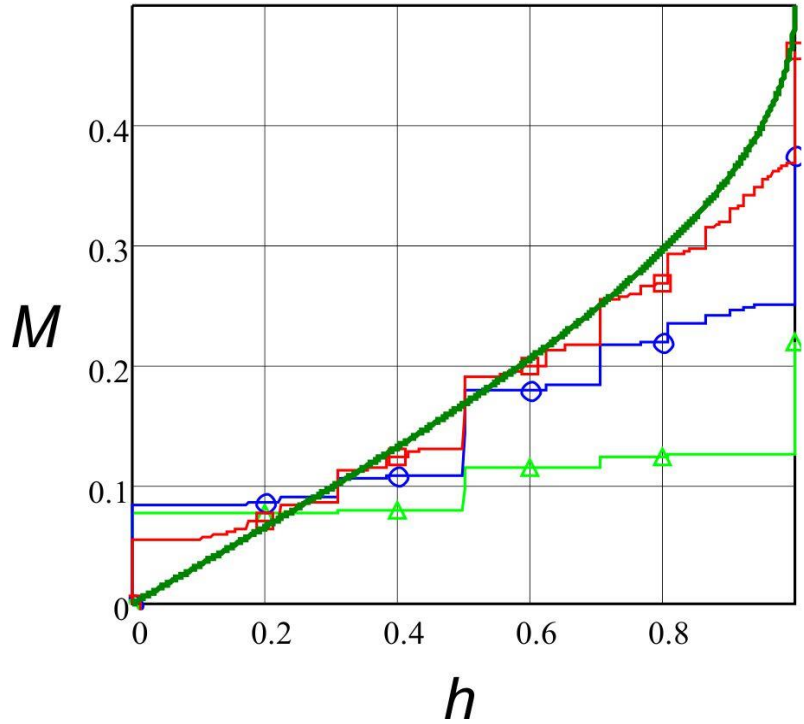

*Fig. 2. Field dependencies of magnetization of dilute XX chain at T=0 for p = 0.25 (△), 0.5(o), 0.75(□), 1(thick solid line). The sum in Eq. (14) is limited by L = 20 terms.*

We can get simple expressions for some largest $M$ jumps near some rational $q_F(h)$. Thus for irrational $q_F(h) = \frac{1}{k} + \delta$, $\delta \to 0$, i. e. for fields $h \approx \cos\frac{\pi}{k} \equiv h_k$, $k = 2,3,4,5,6,...$

$$h_k = 0, \frac{1}{2}, \frac{\sqrt{2}}{2}, \frac{\sqrt{5}+1}{4}, \frac{\sqrt{3}}{2},...$$

we have from Eq. (14) putting for $l > k - 2 \geq 1$, $l + 1 = km + n$ with integer $m \geq 1$ and $n \geq 0$

$$2M \approx (1-p)^2 \left\{ \sum_{l=1}^{k-2} p^l l + \sum_{m=1}^{\infty} p^{km-1} \sum_{n=0}^{k-1} p^n \left[ m(k-2) + n - 1 + \delta_{n,0}(1 + sign(h - h_k)) \right] \right\}$$

Fulfilling summation we get near $h_k$ the expression (valid also for $k = 2$)

$$2M \approx p\frac{1 - p^{k-2}}{1 - p^k} + sign(h - h_k) v_k(p) \qquad (16)$$

where $v_k(p)$ is number of magnetic clusters (per site) whose sites' numbers are multiple of $k$,

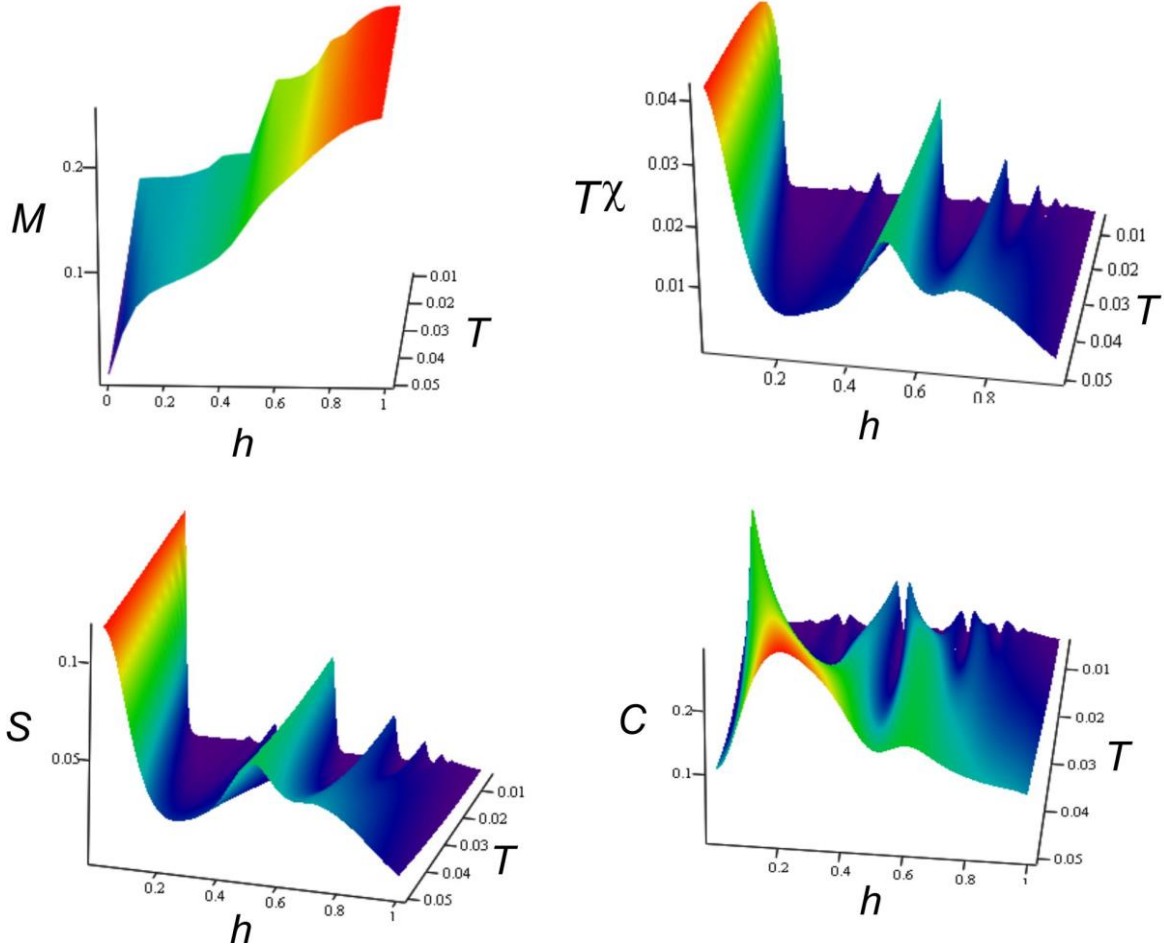

Fig.3. The dependencies of thermodynamic quantities on h < 1 and T << J for the chain with p = 0.5. The sums in Eqs. (8-11) are limited by L = 20 terms.

$$v_k(p) = p^{k-1}\frac{(1-p)^2}{1-p^k} \qquad (17)$$

At $p \to 1$ Eq. (16) gives magnetization of the pure chain at $h_k$ as $M = \frac{k-2}{2k}$ in conformity with Eq. (15).

For lowest $k$ values we have

$$k=2, \ h_2 = 0, \ 2M \approx sign(h)\, p\frac{1-p}{1+p}$$

$$k=3, \ h_3 = \frac{1}{2}, \ 2M \approx \frac{p}{1+p+p^2}\left[1 + sign(h-\frac{1}{2})p(1-p)\right]$$

$$k = 4, \quad h_4 = \frac{\sqrt{2}}{2}, \quad 2M \approx \frac{p}{1+p^2}\left[1 + sign\left(h - \frac{\sqrt{2}}{2}\right)p^2\frac{1-p}{1+p}\right]$$

At $T \to 0$ $S$, $\chi$, $C$ goes to zero for irrational values of $q_F(h)$. However, when field tends to rational value $q_F(h)$ linearly in $T \to 0$, i. e.

$$h = \cos\left(\pi\frac{m}{k}\right) + 2\alpha T / J,$$

with some constant *α* and irreducible fraction *m/k*, then $S$, $T\chi$, $C$ stay finite

$$S = v_k(p)\left[\ln(2\cosh\alpha) - \alpha\tanh\alpha\right], \quad T\chi = \frac{v_k(p)}{4\cosh^2\alpha}, \quad C = v_k(p)\frac{\alpha^2}{\cosh^2\alpha}$$

Here $v_k(p)$ is given by Eq. (17).

Thus the dependencies of thermodynamic quantities on *h* and *T* are represented by surfaces with a number of wrinkles at low *T* as Fig. 3 shows.

### 3. Correlation functions

The static paired correlation functions of spins define the intensity of magnetic diffraction of neutrons on magnetic materials so to know them is important for experimental identification of a model which gives adequate description of real magnet. We consider average two-point correlators

$$C_r^\alpha = \frac{1}{4}\langle\langle\sigma_n^\alpha\sigma_{n+r}^\alpha\rangle_T\rangle_p, \quad \alpha = x, y, z$$

For these correlators there are two possibilities to be non-zero in dilute chain: either sites *n* and *n+r* both belong to the same magnetic cluster or they belong to the different ones. Hence, for $r \geq 3$ we have

$$C_r^\alpha = \frac{1}{4}\sum_{l=r+1}^\infty w_l \sum_{n=1}^{l-r}\langle\sigma_n^\alpha\sigma_{n+r}^\alpha\rangle_{T,l} + \frac{1}{4}\sum_{n=1}^{r-2}\sum_{l=n}^\infty w_l\langle\sigma_n^\alpha\rangle_{T,l}\sum_{n'=1}^{r-1-n}\sum_{l'=n'}^\infty w_{l'}\langle\sigma_{n'}^\alpha\rangle_{T,l'}$$

$$= \sum_{l=r+1}^\infty w_l \sum_{n=1}^{l-r}C_{l,n,r}^\alpha + \delta_{\alpha,z}\sum_{n=1}^{r-2}\sum_{l=n}^\infty w_l M_{l,n}\sum_{n'=1}^{r-1-n}\sum_{l'=n'}^\infty w_{l'}M_{l',n'}$$

(18)

$$w_l = (1-p)^2 p^l, \quad M_{l,n} = \frac{1}{2}\langle\sigma_n^z\rangle_{T,l}, \quad C_{l,n,r}^\alpha = \frac{1}{4}\langle\sigma_n^\alpha\sigma_{n+r}^\alpha\rangle_{T,l}$$

Here $\langle...\rangle_{T,l}$ means the Gibbs average for pure magnetic cluster with *l* sites and $w_l$ is the probability that a given site belongs to such cluster. Note that due to the symmetry between two cluster's edges

$$M_{l,n} = M_{l,l-n+1}, \quad C^{\alpha}_{l,n,r} = C^{\alpha}_{l,l-n-r+1,r}.$$

For *r* = 2 there is just one empty site between different clusters so

$$C^{\alpha}_2 = \frac{1}{4}\left\langle\left\langle\sigma^{\alpha}_i \sigma^{\alpha}_{i+2}\right\rangle_T\right\rangle_p = \sum_{l=3}^{\infty} w_l \sum_{n=1}^{l-2} C^{\alpha}_{l,n,2} + \delta_{\alpha,z}(1-p)^{-1}\left(\sum_{l=1}^{\infty} w_l M_{l,1}\right)^2 \quad (19)$$

and for *r* =1 there no contribution from different clusters,

$$C^{\alpha}_1 = \frac{1}{4}\left\langle\left\langle\sigma^{\alpha}_i \sigma^{\alpha}_{i+1}\right\rangle_T\right\rangle_p = \sum_{l=2}^{\infty} w_l \sum_{n=1}^{l-1} C^{\alpha}_{l,n,1}. \quad (20)$$

Thus we should have the expressions for local magnetizations $M_{l,n}$ and correlators $C^{\alpha}_{l,n,r}$ in finite clusters. So we get

$$M_{l,n} = \frac{1}{2}\langle\sigma^z_n\rangle_{T,l} = \langle a^+_n a_n\rangle_{T,l} - \frac{1}{2} = \frac{1}{l+1}\sum_{k=1}^{l}\tanh u_{k,l} \sin^2\left(\pi\frac{kn}{l+1}\right) \quad (21)$$

$$C^z_{l,n,r} = \frac{1}{4}\langle\sigma^z_n \sigma^z_{n+r}\rangle_{T,l} = M_{l,n} M_{l,n+r} - D^2_{l,n,r}, \quad r \neq 0 \quad (22)$$

$$D_{l,n,r} \equiv \langle a^+_n a_{n+r}\rangle_{T,l} = \frac{1}{l+1}\sum_{k=1}^{l}\tanh u_{k,l} \sin\frac{\pi k n}{l+1}\sin\frac{\pi k(n+r)}{l+1}, \quad r\neq 0, \; D_{l,n,0} = M_{l,n} + \frac{1}{2}. \quad (23)$$

The expression for $C^x_{l,n,r} = C^y_{l,n,r} \equiv C^{\perp}_{l,n,r}$ is more cumbersome involving determinant of $r \times r$ matrix [1] which in our notations reads

$$C^{\perp}_{l,n,r} = \frac{1}{4}\begin{vmatrix} G_{l,n,1} & G_{l,n,2} & \cdots & G_{l,n,r} \\ G_{l,n+1,0} & G_{l,n+1,1} & \cdots & G_{l,n+1,r-1} \\ \vdots & \vdots & \ddots & \vdots \\ G_{l,n+r-1,2-r} & G_{l,n+r-1,3-r} & \cdots & G_{l,n+r-1,1} \end{vmatrix}, \quad G_{l,n,r} = 2D_{l,n,r} - \delta_{r,0} \quad (24)$$

Eqs. (18) - (24) allow to obtain the average correlators with any prescribed precision.

Equations (18-20) can be simplified in case of weak dilution, $1 - p \ll 1$, when major contributions come from large clusters with $l \gg 1$ so we can put

$$M_{l,n} \approx M_\infty \equiv \lim_{n,l\to\infty} M_{l,n} = \int_0^\pi \frac{d\vartheta}{2\pi} \tanh \frac{J}{2T}(h-\cos\vartheta), \tag{25}$$

$$C_{l,n,r}^\alpha \approx C_{\infty,r}^\alpha \equiv \lim_{n,l\to\infty} C_{l,n,r}^\alpha,$$

$$C_{\infty,r}^z = M_\infty^2 - D_{\infty,r}^2, \quad r \neq 0$$

$$D_{\infty,r} \equiv \lim_{n,l\to\infty} D_{l,n,r} = \int_0^\pi \frac{d\vartheta}{2\pi} \cos(\vartheta r) \tanh \frac{J}{2T}(h-\cos\vartheta), r \neq 0, \quad D_{\infty,0} = M_\infty + \frac{1}{2}. \tag{26}$$

$$C_{\infty,r}^\perp = \frac{1}{4} \begin{vmatrix} G_1 & G_2 & \cdots & G_r \\ G_0 & G_1 & \cdots & G_{r-1} \\ \vdots & \vdots & \ddots & \vdots \\ G_{2-r} & G_{3-r} & \cdots & G_1 \end{vmatrix}, \quad G_r = 2D_{\infty,r} - \delta_{r,0}, \tag{27}$$

to obtain

$$C_r^\alpha \approx p^{r+1} C_{\infty,r}^\alpha + \delta_{\alpha,z} M_\infty^2 p^2 \left[1 - (r-1)p^{r-2} + (r-2)p^{r-1}\right], r > 2 \tag{28}$$

$$C_2^\alpha \approx p^3 C_{\infty,2}^\alpha + \delta_{\alpha,z} M_\infty^2 p^2 (1-p), \quad C_1^\alpha \approx p^2 C_{\infty,1}^\alpha \tag{29}$$

Thus in case of weak dilution the correlations are determined essentially by those of pure infinite chain. At $T = 0$ Eqs. (25, 26) again give simple known expressions [2-4]

$$M_\infty = \frac{1}{2} - q_F, \quad D_{\infty,r} = -\frac{\sin \pi q_F r}{\pi r}, \quad r \neq 0$$

which allow to find $C_{\infty,r}^z = M_\infty^2 - D_{\infty,r}^2$ [4] and asymptotic of $C_{\infty,r}^\perp$ at $h = 0$ [8-10]

$$C_{\infty,r}^\perp \approx c(-1)^r / \sqrt{r}, \quad c = 0.147088.$$

Our numerical simulations show that in finite fields

$$C_{\infty,r}^\perp \approx A(h)(-1)^r / \sqrt{r}, \quad A(h) = c(1-h^2)^{1/4}, \quad r \to \infty.$$

Thus in weak dilution regime the $C_r^\perp$ power-law asymptotic is modified by the exponential prefactor $p^{r+1}$ while $C_r^z \approx p^2 M_\infty^2 \left[1-(1-p)rp^{r-2}\right]$ at large $r$.

In general, at $T = 0$ we can perform summation in Eqs. (21, 23) with the result

$$M_{l,n} = \frac{1}{2(l+1)} \left[ f(l, m_l^F, 2n) + l - 2m_l^F + \delta_{m_l^F, q_F(l+1)}(1 - \cos 2\pi q_F n) \right] \tag{30}$$

$$D_{l,n,r} = \frac{1}{2(l+1)}\left[\begin{array}{l} f(l,m_l^F,2n+r)-f(l,m_l^F,r) \\ +\delta_{m_l^F,q_F(l+1)}\left[\cos\pi q_F r - \cos\pi q_F(2n+r)\right]\end{array}\right], r\neq 0, D_{l,n,0}=M_{l,n}+\frac{1}{2} \quad (31)$$

$$f(l,m,r) = \sin\frac{\pi r(2m+1)}{2(l+1)} / \sin\frac{\pi r}{2(l+1)}. \quad (32)$$

These compact expressions make possible to obtain numerically the average correlators at $T=0$ with reasonable precision. Figures 4, 5 show the distance dependence of the correlators $C_r^\perp$ and $C_r^z$ at $T=0$, $h=0.5$. The data on these figures are obtained by cutting the $l$-sums in Eq. (18) - (20) at $L=15$ for $C_r^\perp$ and at $L=30$ for $C_r^z$. The relative error $\varepsilon$ we get dropping the terms with $l > L > r$ in these equations can be estimated as in Eq. (13)

$$\varepsilon \approx p^{L-r}\frac{L(1-p)+1}{r(1-p)+1}$$

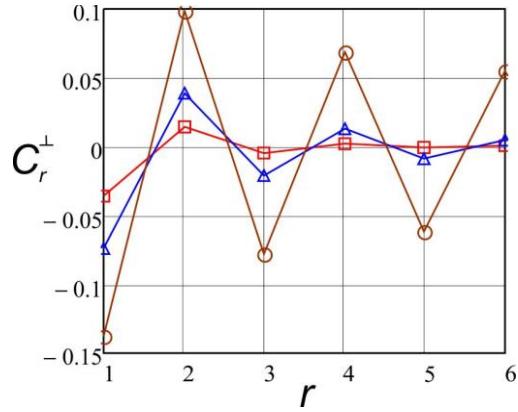

Fig.4. $C_r^\perp$ at $T=0$ and $h=0.5$ for $p=0.5$ (□), 0.75 (Δ), 1 (o). The $l$-sums are cut at $L=15$.

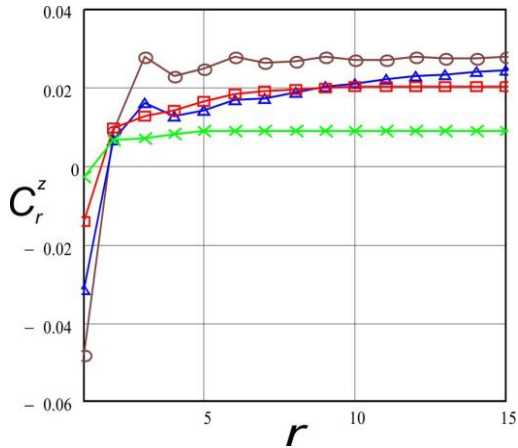

Fig. 5. $C_r^z$ at $T=0$ and $h=0.5$ for $p=0.25$ (x), 0.5 (□), 0.75 (Δ), 1 (o). The $l$-sums are cut at $L=30$.

.

We can see that under dilution the oscillations die out fast in $C_r^\perp$ at large $r$ owing to the exponentially small probability (less than $p^r$) to find the magnetic clusters larger than $r$. Similarly, $C_r^z$ tends fast to a constant proportional to $p^2 M_\infty^2$.

Figures 6, 7 show the numerical results for the field dependencies of nearest-neighbor correlators at $T = 0$ exhibiting the jumps at rational $q_F$ values. These data are compared with correlators of pure chain $C_{\infty,1}^\perp = -\sqrt{1-h^2}/2\pi$ and $C_{\infty,1}^z = M_\infty^2 - (1-h^2)/\pi^2$. Note that $|C_1^\perp|$ steadily diminishes with diminishing of $p$, but $|C_1^z|$ can be greater in dilute chain than in the pure one. The latter effect correlates with that found for the magnetization, cf. Fig. 2.

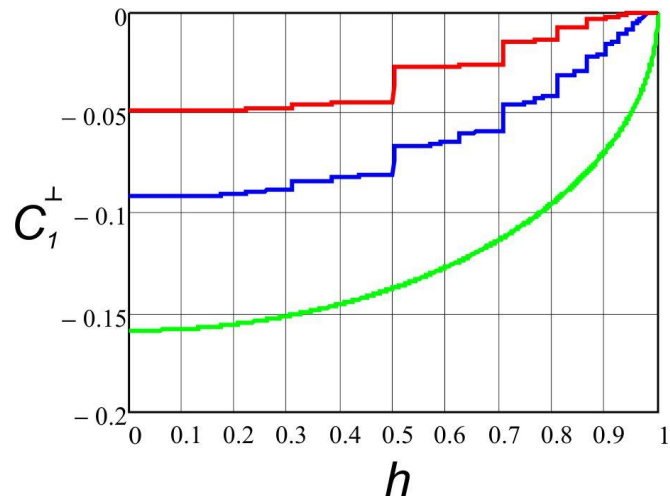

Fig.6. The field dependencies of $C_1^\perp$ for $p = 0.5, 0.75, 1$ (from top to bottom). The l-sums are cut at L=15.

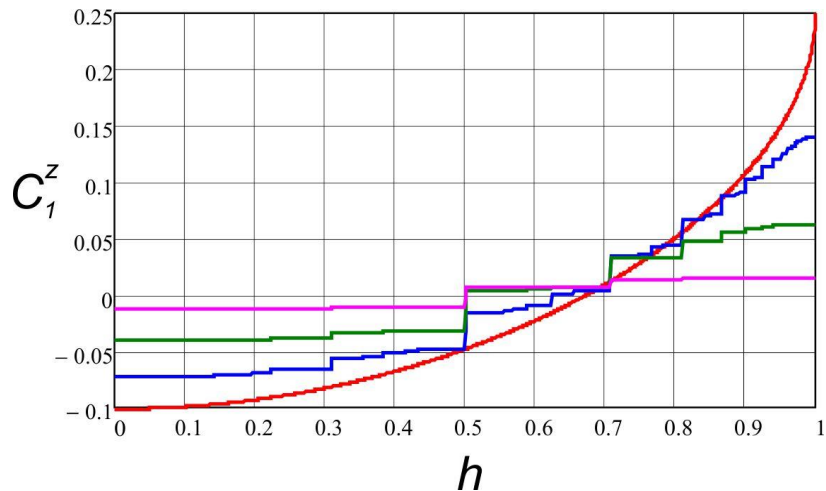

Fig.7. The field dependencies of $C_1^z$ $p = 1, 0.75, 0.5, 0.25$ (from top to bottom in r.h.s. of the figure). The l-sums are cut at $L = 30$.

## 4. Conclusions

The present results show that dilute isotropic XY chain in transverse magnetic field is the rare macroscopic object where the signatures of numerous quantum phase transitions can be observed in macroscopic experiments at low $T$. Actually, with due precision, one can find arbitrary large number of quantum transitions in every finite field's interval at $h < 1$. Moreover, these transitions allow for exact analytical description of its mechanism as the change of the ground state (level-crossing) under variation of external field. The continuous level-crossing is inherent property of many others quantum phases with permanent criticality - the gapless (algebraic) spin-liquid states which are shown to exist in quantum spin chains [12], ladders [13] and planar models [14, 15]. The dilution of such chains and ladders would result in the discretization of their spectra thus inducing the quantum jumps under variation of couplings and field similar to those of XX chain. In the gapless phases of planar models such jumps could also exist but should be much smaller owing to the great variety of possible magnetic clusters.

The present results can be used for some practical aids. Thus if one suspects that some crystal with the uncoupled chains of magnetic ions of one-half spins in its structure can be described by the XX model it suffices to dilute these chains to verify the appearance of highly nonlinear behavior of the ordinary thermodynamic parameters, cf. Fig. 3. Also the dilution of such crystals can be used to enhance their magnetization and longitudinal correlations due to peculiar quantum finite-size effects, see Figs. 2, 7.

## Acknowledgments

The useful discussions with G.Y. Chitov are gratefully acknowledged. We acknowledge support from the Southern Federal University, grant # 213.01-2014/011-ВГ.